\documentstyle[erice98,12pt,psfig]{article}
\begin{document}
\title{Nonlocal Kinetic Equation and Simulations of Heavy Ion Reactions}
\author{Klaus Morawetz}
\address{Fachbereich Physik, University Rostock,
D-18055 Rostock, Germany}
\author{Pavel Lipavsk\'y and V\'aclav \v Spi\v cka}
\address{Institute of Physics, Academy of Sciences, Cukrovarnick\'a 10,
16200 Praha 6, Czech Republic}
\maketitle

\abstracts{A kinetic equation which combines the quasiparticle drift 
of Landau's equation with a dissipation governed by a nonlocal and 
noninstantaneous scattering integral in the spirit of Enskog 
corrections is discussed. Numerical values of the off-shell 
contribution to the Wigner distribution, of the collision duration
and of the collision nonlocality are presented for different realistic 
potentials. On preliminary results
we show that simulations of quantum molecular dynamics extended by the 
nonlocal treatment of collisions leads to a broader proton distribution 
bringing the theoretical spectra closer towards the experimental values 
than the local approach.} 

\section{Introduction}

One of the long standing problems in nuclear physics is to find the 
equation of state of nuclear matter \cite{SG86}. In the absence of any 
direct measurement, it is hoped that the equation of state can be 
deduced from heavy ion reactions via dynamical simulation of the 
fragmentation scenario. Most simulations rely on the local and 
instantaneous treatment of binary collisions as they appear in 
the Boltzmann equation  
\begin{eqnarray}
{\partial f_1\over\partial t}+{\partial\varepsilon_1\over\partial k}
{\partial f_1\over\partial r}-{\partial\varepsilon_1\over\partial r}
{\partial f_1\over\partial k}
&=&\sum_b\int{dpdq\over(2\pi)^5}
\delta\left(\varepsilon_1+\varepsilon_2-
\varepsilon_3-\varepsilon_4\right)
\left|T_{ab}\left(\varepsilon_1+\varepsilon_2,k,p,q,t,r\right)\right|^2
\nonumber\\
&&\times
\Bigl[f_3f_4\bigl(1-f_1\bigr)\bigl(1-f_2\bigr)-
\bigl(1-f_3\bigr)\bigl(1-f_4\bigr)f_1f_2\Bigr].
\label{1}
\end{eqnarray}
The arguments of distributions $f$ and energies $\varepsilon$ are 
shortened as $f_1\equiv f_a(k,r,t)$, $f_2\equiv f_b(p,r,t)$,
$f_3\equiv f_a(k-q,r,t)$, and $f_4\equiv f_b(p+q,r,t)$, with momenta
$k,p,q$, coordinate $r$, time $t$, and spin and isospin $a,b$. The
local picture of the collision is reflected by the same coordinate 
$r$ at all distributions, the instantaneous by the same time $t$.

\begin{figure}
\parbox[h]{14.5cm}{
\parbox[]{8cm}{
  \psfig{figure=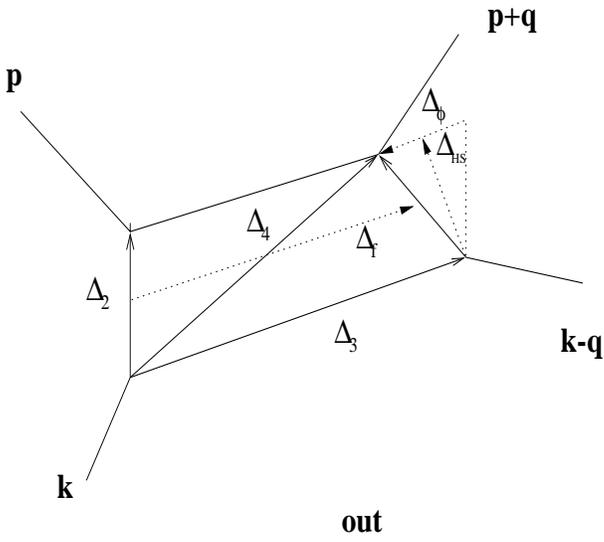,width=8cm,height=7cm}}
\hspace{1cm}
\parbox[]{5cm}{\vspace{5cm}\caption{A nonlocal binary collision.
\label{soft}}}
}
\end{figure}

A real binary collision is neither local nor instantaneous. A nonlocal
picture of a collision is schematically drawn in figure~\ref{soft}.
Let us introduce individual nonlocal corrections step by step for
simple but useful models. The first nonlocal corrections have been 
introduced by Enskog for the classical gas of hard spheres \cite{CC90}. 
The collision of hard spheres is instantaneous so that each trajectory
is broken only at a single point.
Accordingly, in figure~\ref{soft} 
$\Delta_3=0$ and $\Delta_4-\Delta_2=0$, i.e., 
$\Delta_{\rm f}=0$ and $\Delta_\phi=0$. At the
instant of the collision, the particles are displaced by the sum of
their radii in the direction of the transferred momentum what is
described by a nonzero vector $\Delta_{\rm HS}$. Corresponding changes
in the kinetic equation enter the position of the ongoing particle,
$f_2=f_b(p,r-\Delta_{\rm HS},t)$ and $f_4=f_b(p+q,r-\Delta_{\rm HS},t)$
while other arguments remain unchanged. The equation of state evaluated 
from the kinetic equation with the nonlocal scattering integral is of 
the van der Waals type covering the excluded volume \cite{CC90,HCB64}. 
For nuclear matter, Enskog's corrections have been first discussed by 
Malfliet \cite{M84} and recently implemented by Kortemayer, Daffin and 
Bauer \cite{KDB96}.

Another simple model one obtains assuming that colliding particles form
an unstable molecule of an average time of life $\Delta_t$ called the 
collision duration or the collision delay. Neglecting the size of the
molecule, $\Delta_2=0$ and $\Delta_4-\Delta_3=0$, i.e., $\Delta_{\rm HS}
=0$ and $\Delta_\phi=0$, the only nonzero displacement $\Delta_{\rm f}$ 
measures a distance between points where molecule forms and breaks up. 
This distance is given by ${\Delta_{\rm f}}={\Delta_t} v^{\rm mol}$ with 
the molecular velocity $v^{\rm mol}=(k+p)/(m_a+m_b)$. Corresponding 
changes in the kinetic equation enter the time argument and positions 
of final states, $f_3=f_a(k-q,r-\Delta_{\rm f},t-\Delta_t)$ and $f_4=
f_b(p+q,r-\Delta_{\rm f},t-\Delta_t)$. In the equation of state, the 
collision duration results in the same kind of terms as the presence of 
stable molecules. The finite duration of nucleon-nucleon collisions and 
its thermodynamic consequences has been studied by Schmidt, R\"opke and
Schulz \cite{SRS90}, its effect on the pressure has been discussed only 
recently by Danielewicz and Pratt \cite{DP96}. The noninstantaneous 
scattering integral and its consequences for the linear response has 
been first discussed for electrons in semiconductors scattered by 
resonant levels \cite{SLMab96}.

In a real collision, the two particles keep a finite distance, 
$\Delta_{\rm HS}\not=0$, and interact for a finite time, $\Delta_t\not=
0$ and $\Delta_{\rm f}\not=0$. Moreover, particles rotate one against
the other as described by the rotation displacement $\Delta_{\phi}$.
Consequently we obtain the following scenario of Fig. \ref{soft}: Two 
particles approach until they reach a distance $\Delta_2$. Then they 
form a molecule living for $\Delta_t$ and traveling over a distance 
$\Delta_{\rm f}$. During this propagation the molecule rotate as given 
by $\Delta_\phi$. Collecting all three shifts we obtain the nonlocal 
and noninstantaneous kinetic equation (\ref{9}) derived in 
\cite{SLM96,M98} with the help of the method introduced in \cite{SLMab96}. 
The resulting arguments of the kinetic equation read finally 
$f_1=f_a(k,r,t)$, $f_2=f_b(p,r-\Delta_2,t)$, 
$f_3=f_a(k-q,r-\Delta_3,t-\Delta_t)$ and 
$f_4=f_b(p+q,r-\Delta_4,t-\Delta_t)$.

While the above microscopic picture of nonlocal and noninstantaneous
isolated collisions is intuitively clear, it is less transparent how to
define the same corrections for quasiparticles which carry a part of
the interaction in the quasiparticle reconstruction of their energies
and wave function norms. This question requires a systematic approach 
as it was first presented in \cite{SLM96}. This derivation follows 
Baerwinkel \cite{B69} in starting from nonequilibrium Green's functions 
and keeping all gradient contributions to the scattering integral, but 
instead of the quasiparticle approximation, the extended quasiparticle 
approximation is used. Here we use numerical results to discuss the key
steps and consequences of this approach.

\section{Kinetic equation}
We start our derivation of the kinetic equation from the quantum
transport equation for the nonequilibrium Green's functions first 
obtained by Kadanoff and Baym, see \cite{D84},
\begin{equation}
-i\left[G^{-1}_0-{\rm Re}\Sigma,G^<\right]-
i\left[{\rm Re}G,\Sigma^<\right]={1\over 2}\left\{G^>,\Sigma^<\right\}-
{1\over 2}\left\{G^<,\Sigma^>\right\},
\label{2}
\end{equation}
where $[,]$ and $\{,\}$ denote commutators and anticommutators, ${\rm
Re}G={1\over 2}(G^R+G^A)$ is the hermitian part of the propagator. The
center of interest is the particle correlation function $G^<(1,2)=
{\rm TR}\left(\hat\rho\Psi^\dagger(2)\Psi(1)\right)$. Its time
evolution, however, requires to know the accessible states given by the
hole correlation function $G^>(1,2)={\rm TR}\left(\hat\rho\Psi(1)
\Psi^\dagger(2)\right)$, and a dynamics of interactions specified by
the selfenergy $\Sigma$. Individual terms in (\ref{2}) have specific
physical content. The $G_0^{-1}$ describes a free motion of particles 
and is renormalized by ${\rm Re}\Sigma$. The ${\rm Re}G$ describes the
off-shell motion after the collision. The first and second 
anticommutators are the scattering-in and -out. 

The dynamics of interaction reflects selected models and approximations. 
For simplicity we assume that protons and neutrons are of equal mass $m$, 
interact via an instant potential $V$, and there is no spin-flipping 
mechanism. As common, the self-energy is constructed from retarded and 
advanced two-particle T-matrices $T^{R,A}$ in the Bethe-Goldstone 
approximation \cite{D84,MR95} as
\begin{equation}
\Sigma^<(1,2)=T^R(1\bar 3,\bar 5\bar 6)T^A(\bar 7\bar 8,2\bar 4)
G^>(\bar 4,\bar 3)G^<(\bar 5,\bar 7)G^<(\bar 6,\bar 8),
\label{3}
\end{equation}
and $\Sigma^>$ is obtained from (\ref{3}) by an interchange
$>\leftrightarrow <$. Numbers are cumulative variables, $1\equiv 
(t_1,r_1,a_1)$, and bars denote internal variables that are integrated 
over. Missing commas in arguments signal that the time arguments are 
identical, e.g., $t_{\bar 3}=t_1$ and $t_{\bar 5}=t_{\bar 6}$, the 
T-matrices are thus double-time functions. 
The approximations of set (\ref{2}-\ref{3}) are specified in the mixed
representation, [off-diagonal elements in spin and isospin are excluded,
$a_1=a_2=a$]
\begin{equation}
G^<(1,2)=\int{d\omega\over 2\pi}{dk\over(2\pi)^3}
{\rm e}^{ik(r_1-r_2)-i\omega(t_1-t_2)}G^<_a
\left(\omega,k,r,t\right)_{r={r_1+r_2\over 2},t={t_1+t_2\over 2}}.
\label{4}
\end{equation}

\subsection{Off-shell motion}
Equations (\ref{2}-\ref{3}) completed with the Dyson equation for $G^R$
and the ladder equation for $T^R$ form a closed set for $G^<$. This set
is converted to equations for the quasiparticle distribution $f$ with
the help of the extended quasiparticle approximation 
\cite{KM93,MR95,SL95,BKKS96} 
\begin{equation}
G^{\stackrel{>}{<}}_{1,\omega}=
\left(\!\begin{array}{c}1\!-\!f_1\\ f_1\end{array}\!\right)
2\pi z_1\delta(\omega-\varepsilon_1)+
{\rm Re}{\Sigma^{\stackrel{>}{<}}_{1,\omega}
\over(\omega-\varepsilon_1)^2},
\label{5}
\end{equation}
where $G_{1,\omega}\equiv G_a(\omega,k,r,t)$ and similarly $\Sigma$. 
The first term is singular and provides the dominant quasiparticle
contribution on the energy shell. The second term is regular and 
contributes out of the energy shell. The approximative form of this
off-shell contribution is consistent with the lowest approximation of
the wave-function renormalization, $z_1=1+\left.
{\partial\over\partial\omega}{\rm Re}\Sigma_{1\omega}\right|_
{\varepsilon_1}$.
\begin{figure}[h]
\parbox[h]{18.2cm}{
\parbox[]{8cm}{
  \psfig{figure=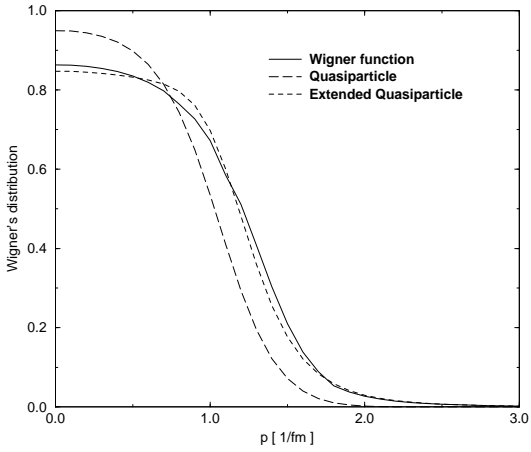,width=7cm,height=6cm,angle=-90}}
\hspace{2cm}
\parbox[]{8cm}{
\caption{The equilibrium occupation of momentum states given by Wigner's 
distribution $\rho$ (full line) is compared with the quasiparticle 
distribution $f$ (long dashed line) and with the extended quasiparticle 
approximation (short dashed line) at the temperature $T=10$~MeV.
\label{if1} } }}
\end{figure}
Before we employ the extended quasiparticle approximation (\ref{5}) to
convert the transport equation (\ref{2}) into a kinetic equation, it is 
useful demonstrate its physical content. The amplitude of the off-shell
contributions and the need to treat them separately as in (\ref{5}) can
be seen in figure~\ref{if1} where we compare the Wigner distribution
$\rho_1={1\over 2\pi}\int d\omega G^<_{1,\omega}$ obtained from the
exact equilibrium correlation functions $G^<$ with its extended
quasiparticle approximation (\ref{5}) and the plain quasiparticle
approximation $G^<_{1,\omega}=f_1 2\pi\delta(\omega-\varepsilon_1)$
which yields the Fermi-Dirac distribution. We note that the simple 
Yamaguchi potential \cite{Y59} has been used in the T-matrix
for this comparison. One can 
see that the off-shell contribution given by the difference between the
Wigner and the Fermi-Dirac distributions is not small, in particular at 
the high momenta region where the power-law off-shell tails always 
dominate over the exponentially falling quasiparticle distribution. 
Formula (\ref{5}) provides inevitable and sufficiently precise off-shell 
corrections. 

Out of equilibrium one has to evaluate $\Sigma^<$ which is similar to 
an evaluation of the scattering integral. From the extended quasiparticle 
approximation one then obtains the high-momenta tails found by Danielewicz
\cite{D841} and K\"ohler \cite{HSK96} in direct numerical treatments of
non-equilibrium Green's functions.

\subsection{Quasiclassical trajectories}
Now we substitute (\ref{5}) into (\ref{2}) and neglect all gradient 
terms but linear. Within nuclear physics, the gradient expansion of the 
self-energy (\ref{3}) is commonly omitted since it is a tedious task. It
results in one nongradient and nineteen gradient terms that are
analogous to those found within the chemical physics \cite{NTL91,H90}.
All these terms can be recollected into a nonlocal and noninstantaneous
scattering integral that has an intuitively appealing structure of a
nonlocal Boltzmann equation (\ref{1}) 
\begin{eqnarray}
{\partial f_1\over\partial t}+{\partial\varepsilon_1\over\partial k}
{\partial f_1\over\partial r}-{\partial\varepsilon_1\over\partial r}
{\partial f_1\over\partial k}
&=&\sum_b\int{dpdq\over(2\pi)^5}\delta\left(\varepsilon_1+
\varepsilon_2-\varepsilon_3-\varepsilon_4+2\Delta_E\right)
\Biggl(1-{1\over 2}{\partial\Delta_2\over\partial r}-
{\partial\bar\varepsilon_2\over\partial r}
{\partial\Delta_2\over\partial\omega}\Biggr)
\nonumber\\
&\times &z_1z_2z_3z_4
\left|T_{ab}\!\left(\varepsilon_1\!+\!\varepsilon_2\!-\!
\Delta_E,k\!-\!{\Delta_K\over 2},p\!-\!{\Delta_K\over 2},
q,r\!-\!\Delta_r,t\!-\!{\Delta_t\over 2}\!\right)\right|^2
\nonumber\\
&\times &\Bigl[f_3f_4\bigl(1-f_1\bigr)\bigl(1-f_2\bigr)-
\bigl(1-f_3\bigr)\bigl(1-f_4\bigr)f_1f_2\Bigr],
\label{9}
\end{eqnarray}
with Enskog-type shifts of arguments \cite{SLM96,M98}:
$f_1\equiv f_a(k,r,t)$, $f_2\equiv f_b(p,r\!-\!\Delta_2,t)$,
$f_3\equiv f_a(k\!-\!q\!-\!\Delta_K,r\!-\!\Delta_3,t\!-\!\Delta_t)$, and
$f_4\equiv f_b(p\!+\!q\!-\!\Delta_K,r\!-\!\Delta_4,t\!-\!\Delta_t)$.
In agreement with \cite{NTL91,H90}, all gradient corrections result
proportional to derivatives of the scattering phase shift
\mbox{$\phi={\rm Im\ ln}T^R_{ab}(\Omega,k,p,q,t,r)$},
\begin{equation}
\begin{array}{lclrclrcl}
\Delta_2&=&
{\displaystyle\left({\partial\phi\over\partial p}-
{\partial\phi\over\partial q}-{\partial\phi\over\partial k}
\right)_{\varepsilon_3+\varepsilon_4}}&\ \
\Delta_3&=&
{\displaystyle\left.-{\partial\phi\over\partial k}
\right|_{\varepsilon_3+\varepsilon_4}}&\ \
\Delta_4&=&
{\displaystyle-\left({\partial\phi\over\partial k}+
{\partial\phi\over\partial q}\right)_{\varepsilon_3+\varepsilon_4}}
\\ &&&&&&&&\\
\Delta_t&=&
{\displaystyle \left.{\partial\phi\over\partial\Omega}
\right|_{\varepsilon_3+\varepsilon_4}}&\ \
\Delta_E&=&
{\displaystyle \left.-{1\over 2}{\partial\phi\over\partial t}
\right|_{\varepsilon_3+\varepsilon_4}}&\ \
\Delta_K&=&
{\displaystyle \left.{1\over 2}{\partial\phi\over\partial r}
\right|_{\varepsilon_3+\varepsilon_4}},
\end{array}
\label{8}
\end{equation}
and $\Delta_r={1\over 4}(\Delta_2+\Delta_3+\Delta_4)$. After derivatives, 
$\Delta$'s are evaluated at the energy shell $\Omega\to\varepsilon_3+
\varepsilon_4$. 

For the purpose of discussion, it is advantageous to link the quantum 
displacements (\ref{8}) to intuitively more appealing hard-sphere and 
rotation shifts by relations obvious from figure \ref{soft}
\begin{equation}
\Delta_{\rm HS}=\frac 1 2 (\Delta_4-\Delta_3+\Delta_2),
\ \ \ \ \ \ \ \ \ \ \ \ \ \ \ \ \ \ \ \ 
\Delta_{\phi}= \frac 1 2 (\Delta_4-\Delta_3-\Delta_2).
\label{para}
\end{equation}
For the collision of two isolated nucleons, it is possible to show
that $\Delta_{\rm HS}$ points in the direction of the transferred
momentum $q$. Similarly it follows that the rotation shift is 
orthogonal to $\Delta_{\rm HS}$ and stays in the collision plane.

\begin{figure}
\parbox[h]{20.2cm}{
\parbox[]{10cm}{
\psfig{file=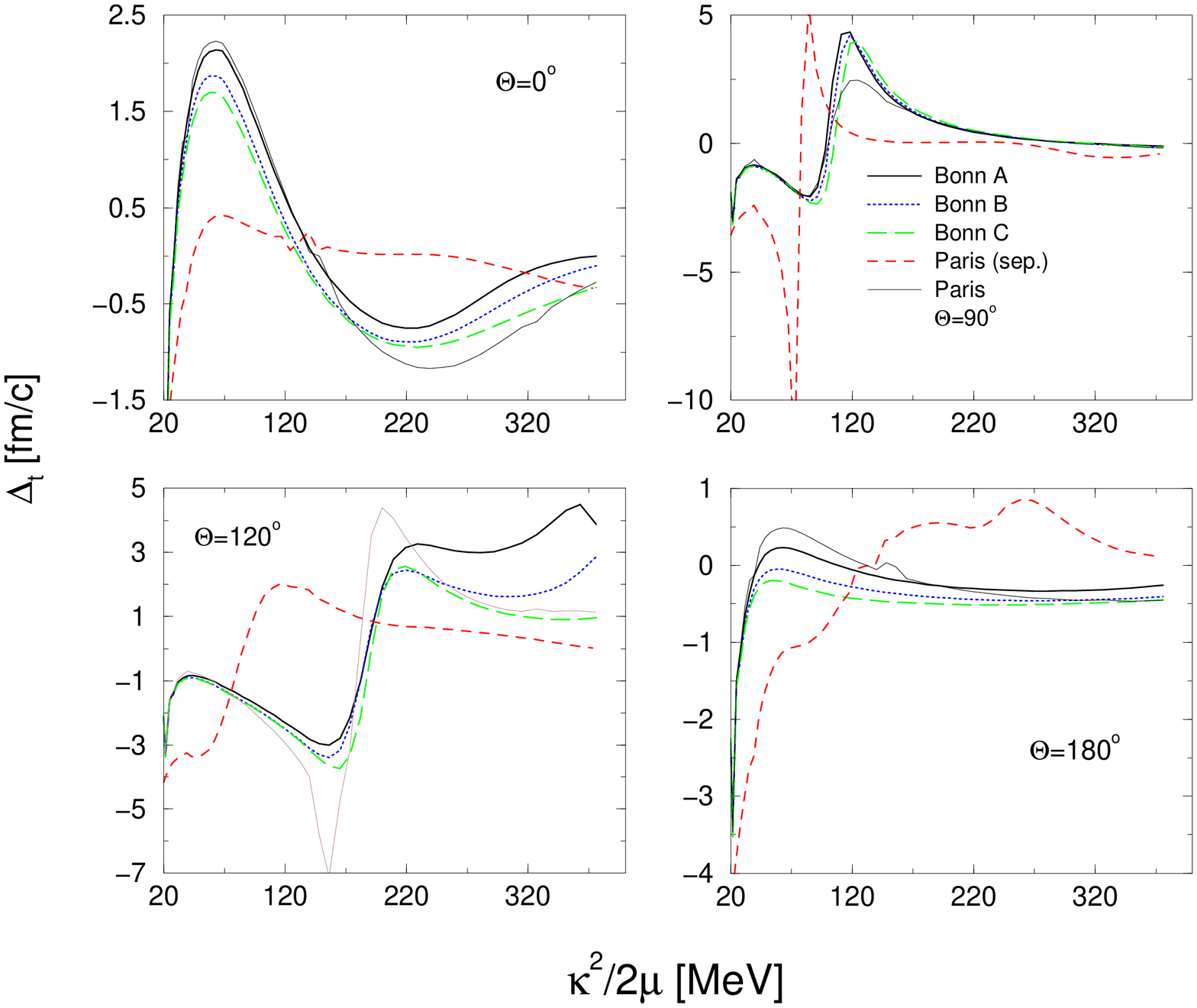,width=9cm,height=10cm,angle=-90}}
\parbox[]{10cm}{
\psfig{file=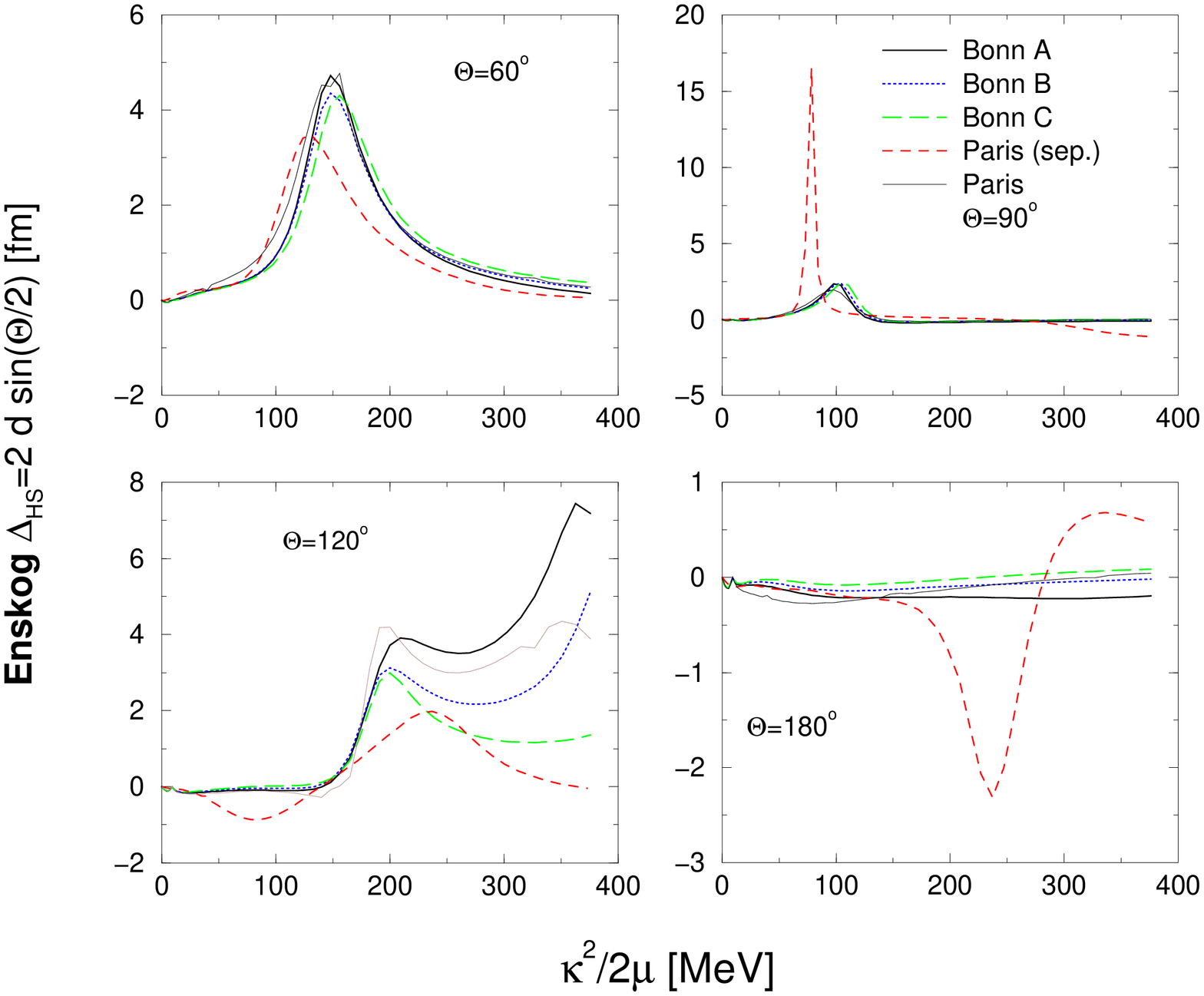,width=9cm,height=10cm,angle=-90}}
}
\caption{The collision delay $\Delta_t$ (left) and the hard core shift 
$|\Delta_{\rm HS}|$ (right) for isolated nucleons in the barycentric 
coordinate system for different deflection angles $\theta$ and 
interaction potentials. 
\label{dt}}
\end{figure}

In figure \ref{dt} we plot the delay time and the amplitude of the
hard-sphere shift for different deflection angles versus lab energy. 
The T-matrix is evaluated with different potentials, Bonn (A-C)
\cite{LMK93}, Paris \cite{LLRVCPT80} and separable Paris \cite{HP84}, 
concerning partial wave coupling up to D-waves \cite{MLSK98}. 
The forward angle 
delay time has a negative minimum at small energies indicating an 
attractive behavior. For very small energies the delay time is rapidly 
decreasing to high negative values reflecting the occurrence of weakly 
bound states. The sharp jump for $\theta=90^o$ is caused by a resonant 
character of this scattering channel. Its value is misplaced and
exaggerated within the separable approximation.
The hard core shift $\Delta_{\rm HS}$ behaves regularly at low energies.
The resonant scattering at $\theta=90^o$ appears as the increase of the
amplitude. Again, the separable potential exaggerates its value. 
Characteristic values $\Delta_t\sim 1$~fm/c and $\Delta_{\rm HS}\sim
1$~fm show that the noninstantaneous and nonlocal treatment of binary
collisions can be important in heavy ion reactions.

\section{Implementation in heavy ion codes}
The selfconsistent evaluation of all $\Delta$'s would be as demanding as
the full Green's function treatment of the system. We employ two kinds
of additional approximations. First, following the approximations used
within the Boltzmann equation, we neglect the medium effect on binary
collision, i.e., use the well known free-space T-matrix. Second, we
rearrange the scattering integral into an instantaneous but nonlocal form.
To eliminate the time delay from distributions we neglect collisions
on the time scale $\Delta_t$ and shift arguments of distributions along
the trajectory of a particle. In the differential form this step derives
as
\begin{eqnarray}
f_a(k-q-\Delta_k,r-\Delta_r^3,t-\Delta_t)
&=&f_a(k-q-\Delta_k,r-\Delta_r^3,t)-{\partial f_3\over\partial t}
\Delta_t
\nonumber\\
&=&f_a\!\left(k-q-\Delta_k-{\partial\varepsilon_3\over\partial r}\Delta_t,
r-\Delta_3+{\partial\varepsilon_3\over\partial k}\Delta_t,t\right).
\label{fd1}
\end{eqnarray}
In the last step we have used the collision-free kinetic equation
${\partial f_3\over\partial t}+{\partial\varepsilon_3\over\partial k}
{\partial f_3\over\partial r}-{\partial\varepsilon_3\over\partial r}
{\partial f_3\over\partial k}=0$. With approximation (\ref{fd1}) and
similar for $f_4$, the scattering integral (\ref{9}) corresponds to
an instantaneous event at time $t$. The collision remains nonlocal 
with modified space displacements of particles at initial states
\begin{equation}
\tilde\Delta_{3,4}=
\Delta_{3,4}-{\partial\varepsilon_{3,4}\over\partial k}\Delta_t.
\label{new}
\end{equation}

The instantaneous approximation brings further simplifications following 
from conservation laws. During the instantaneous process, mean fields 
have no time to pass any momentum and energy to the colliding pair. 
Indeed, assuming the effect on colliding particles only via mean fields, 
from (\ref{8}) one finds $\Delta_k=-{\partial\varepsilon_{3,4}\over
\partial r}\Delta_t$ so that the momentum gain vanishes in (\ref{fd1}).
Similarly, the energy gain $\Delta_E$ vanishes when arguments of
quasiparticle energies in the energy conserving $\delta$ function are
brought to the same time instant. Finally, in agreement with the
continuity of the center of mass motion, one finds that $\Delta_2=
\tilde\Delta_3+\tilde\Delta_4$. The scattering-in thus simplifies as
\begin{equation}
\sum_b\int{dp\over(2\pi)^3}{dq\over(2\pi)^3}
2\pi\delta\left(\varepsilon_1+\varepsilon_2-\varepsilon_3-
\varepsilon_4\right)|T_{ab}|^2\left(k-p,q\right)
f_3f_4\bigl(1-f_1\bigr)\bigl(1-f_2\bigr),
\label{fd3}
\end{equation}
where new arguments of energies and distributions are shifted only in
space, $f_1=f_a(k,r,t)$, $f_2=f_b(p,r-\Delta_2,r,t)$, 
$f_3=f_a(k-q,r-\tilde\Delta_3,t)$ and $f_4=f_b(p+q,r-\tilde\Delta_4,t)$. 
The scattering-out is similar. In the T-matrix we have reduced arguments to 
those which are relevant for free-space collisions. Finally, we would 
like to stress that the amplitudes of displacements are not fitting 
parameters or a matter of an educated guess but evaluated from
the T-matrix, i.e., from an interaction potential.
\vspace{-6ex}
\begin{figure}[h]
\parbox[h]{20.2cm}{
\parbox[]{10cm}{
  \psfig{figure=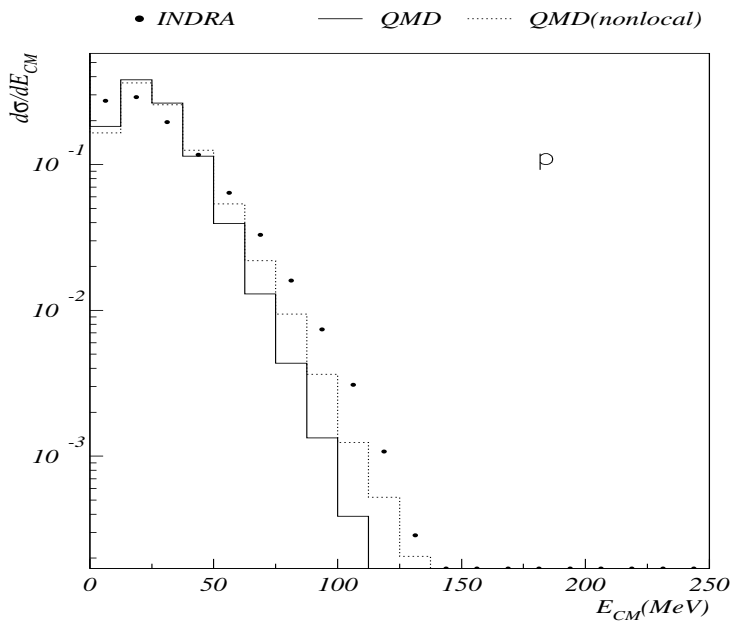,width=10cm,height=12cm}}
\hspace{1mm}
\parbox[]{8cm}{
\vspace{2cm}
\caption{The exclusive QMD proton spectra for central collision of
$^{129}$Xe$\rightarrow ^{119}$Sn at $50$~MeV/A with and without
non-local corrections. The data are extracted from recent INDRA
experiments. The non-local corrections bring the spectrum
towards the experimental values.\label{spek3}}
}
}
\end{figure}
\vspace{-6ex}

In order to investigate the effect of nonlocal collisions in realistic
situations, we have evaluated $\Delta$'s from the separable Paris 
potential \cite{HP84} and implemented the nonlocal scattering
integral of (\ref{fd3}) in a QMD code for the central collision of
$^{129}$Xe$\rightarrow$$^{119}$Sn at $50$~MeV/A.
Figure \ref{spek3} shows the exclusive proton spectra subtracting the 
protons bound in clusters. This procedure is performed within a spanning 
tree model which is known to describe a production of light charged 
cluster in a reasonable agreement with the experimental data. Within 
the local approximation, however, the remaining distribution of 
high-energy protons is too low to meet the experimental values. As one 
can see, the inclusion of nonlocal collisions corrects this shortage 
of the QMD simulation. The increase in the high-energy part follows from
an enhancement of a number of collisions at the pre-equilibrium stage of
the reaction while later stages are not strongly affected \cite{MLSCN98}. 
Accordingly, the production of light clusters is rather insensitive to 
the nonlocal corrections. The improvement of the proton production is 
thus not on cost of worse results in other spectra. 

\section{Thermodynamic properties}

The production of high-energy particles documents a vital role of
nonlocal treatment far from equilibrium. Their role can be best seen on 
thermodynamic observables like density $n_a$ of 
particles $a$, density of energy $\cal E$, and the stress tensor 
${\cal J}_{ij}$ which conserve within the nonlocal and noninstantaneous
kinetic equation (\ref{9}). Integrating (\ref{9}) over momentum $k$ with 
factors $\varepsilon_1$, $k$ and unity one finds \cite{SLM96,M98} that 
each observable has the standard quasiparticle part following from the 
drift
\begin{eqnarray}
{\cal E}^{\rm qp}&=&\sum_a\int{dk\over(2\pi)^3}{k^2\over 2m}f_1+
{1\over 2}\sum_{a,b}\int{dkdp\over(2\pi)^6}
T_{ab}(\varepsilon_1+\varepsilon_2,k,p,0)f_1f_2,
\nonumber\\
{\cal J}_{ij}^{\rm qp}&=&\sum_a\int{dk\over(2\pi)^3}\left(k_j
{\partial\varepsilon_1\over\partial k_i}+
\delta_{ij}\varepsilon_1\right)f_1-\delta_{ij}{\cal E}^{\rm qp},
\nonumber\\
n_a^{\rm qp}&=&\int{dk\over(2\pi)^3}f_1,
\label{10a}
\end{eqnarray}
and the $\Delta$-contribution following from the scattering integral
\begin{eqnarray}
\Delta {\cal E}&=&{1\over 2}\sum_{a,b}\int{dkdpdq\over(2\pi)^9} P
(\varepsilon_1+\varepsilon_2)\Delta_t,
\nonumber\\
\Delta {\cal J}_{ij}&=&{1\over 2}
\sum_{a,b}\int{dkdpdq\over(2\pi)^9} P
\left[(p\!+\!q)_i\Delta_{4j}+(k\!-\!q)_i\Delta_{3j}-p_i\Delta_{2j}\right],
\nonumber\\
\Delta n_a&=&\sum_b\int{dkdpdq\over(2\pi)^9} P \Delta_t,
\label{10}
\end{eqnarray}
where $P=|T_{ab}|^22\pi\delta(\varepsilon_1\!+\!\varepsilon_2\!-
\!\varepsilon_3\!-\!\varepsilon_4)f_1f_2(1\!-\!f_3\!-\!f_4)$. The
arguments denoted by numerical subscripts are identical to those used in
(\ref{1}), for all $\Delta$'s are explicit. 

The density of energy ${\cal E}={\cal E}^{\rm qp}+\Delta{\cal E}$
alternatively results from Kadanoff and Baym formula, ${\cal E}=
\sum_a\int{dk\over(2\pi)^3}\int{d\omega\over 2\pi}
{1\over 2}\left(\omega+{k^2\over 2m}\right)G^<_a(\omega,k,r,t)$,
with $G^<$ in the extended quasiparticle approximation (\ref{5}).
Its complicated form, however, shows that $\cal E$ cannot be easily
inferred from an eventual experimental fit of the kinetic equation as it
has been attempted in \cite{SG86}. The conservation of ${\cal E}$ 
generalizes the result of Bornath, Kremp, Kraeft and Schlanges 
\cite{BKKS96} restricted to non-degenerated systems. The particle 
density $n_a=n_a^{\rm qp}+\Delta n_a$ is also 
obtained from (\ref{5}) via the definition, $n_a=\int{d\omega\over 2\pi}
{dk\over(2\pi)^3}G^<$. This confirms that the extended quasiparticle 
approximation is thermodynamically consistent with the nonlocal and
noninstantaneous corrections to the scattering integral.

For equilibrium distributions, formulas (\ref{10a}) and (\ref{10})
provide equations of state. Two known cases are worth to compare.
First, the particle density $n_a=n_a^{\rm qp}+\Delta n_a$ is identical
to the quantum Beth-Uhlenbeck equation of state \cite{SRS90,MR95,BKKS96},
where $n_a^{\rm qp}$ is called the free density and $\Delta n_a$ the
correlated density. Second, the virial correction to the stress tensor
has a form of the collision flux contribution known in the theory of
moderately dense gases \cite{CC90,HCB64}.

\section{Summary}

In this paper we have discussed the kinetic equation which is consistent 
with thermodynamic observables up to the second order virial coefficient. 
This theory extends the theory of quantum gases and non-ideal plasma to 
degenerated systems. The amplitude of the contribution of the off-shell 
motion is demonstrated on the Wigner distribution which also shows the
precision of the extended quasiparticle approximation. The contribution
of the nonlocal corrections to the scattering integral is documented on
a realistic study of the heavy ion collision within the quantum
molecular dynamics. It should be noted, however, that the separable
Paris potential used in the presented preliminary results is not fully
reliable as seen from comparison of the collision delay and the hard core
displacement. Nevertheless, this preliminary study shows that the nonlocal 
corrections can be evaluated from the T-matrix and incorporated into 
existing Monte Carlo simulation codes.

\section{Acknowledgements}
The authors thank the INDRA collaboration for the use of data prior to
publication. We are grateful to S. Koehler for stimulating discussions, 
Ch. Kuhrts for QMD simulation, A. Schnell for the program of separable 
Paris potential and N.H. Kwong for the other realistic potentials. This work was 
supported from Grant Agency of Czech Republic under contracts 
Nos.~202960098 and 202960021, the BMBF (Germany) under contract 
Nr.~06R0745(0), and the EC Human Capital and Mobility Programme.

\bibliography{delay1,kmsr,kmsr1,kmsr2,kmsr3,kmsr4,kmsr5,kmsr6,kmsr7,spin}
\bibliographystyle{prsty}

\end{document}